\pdfoutput=1
\RequirePackage[fleqn]{amsmath}
\documentclass[12pt]{iopart}

\usepackage{amsfonts}
\usepackage{amssymb}
\usepackage{amsopn}
\usepackage{blindtext}
\usepackage{bm}
\usepackage{setstack}
\usepackage{xfrac}

\usepackage{color}
\DeclareMathOperator*{\argmin}{arg\,min}

\newcommand{\keywords}[1]{\begin{indented}
   \item[]\rm\raggedright Keywords: #1
   \end{indented}}

\begin{document}

\title[Classification of gravitational-wave glitches via dictionary learning]{Classification of gravitational-wave glitches via dictionary learning}

\author{Miquel Llorens-Monteagudo\textsuperscript{1}, Alejandro Torres-Forn\'e\textsuperscript{1,2}, Jos\'e A. Font\textsuperscript{1,3} and Antonio Marquina\textsuperscript{4} }

\address{\textsuperscript{1}Departamento de Astronom\'ia y Astrof\'isica, Universitat de Val\`encia, C/ Dr.~Moliner, 50, Burjassot (Val\`encia E46100, Spain.}
\address{\textsuperscript{2}Max Planck Institute f\"{u}r Gravitationalphysik (Albert Einstein Institute), D-14476 Potsdam-Golm, Germany}
\address{\textsuperscript{3}Observatori Astron\`omic, Universitat de Val\`encia, C/ Catedr\'atico Jos\'e Beltr\'an 2, 46980, Paterna (Val\`encia) E46100, Spain.}
\address{\textsuperscript{4}Departamento de Matem\'aticas, Universitat de Val\`encia, Dr.~Moliner 50, 46100, Burjassot (Val\`encia) E46100, Spain.}
\vspace{10pt}

\begin{abstract}
We present a new method for the classification of transient noise signals (or glitches) in advanced gravitational-wave interferometers. The method uses learned dictionaries (a supervised machine learning algorithm) for signal denoising, and untrained dictionaries for the final sparse reconstruction and classification. We use a data set of 3000 simulated glitches of three different waveform morphologies, comprising 1000 glitches per morphology. These data are embedded in non-white Gaussian noise to simulate the background noise of advanced LIGO in its broadband configuration. Our classification method yields a 96\% accuracy for a large range of initial parameters, showing that learned dictionaries are an interesting approach for glitch classification. This work constitutes a preliminary step before assessing the performance of dictionary-learning methods with actual detector glitches.
\end{abstract}



\section{Introduction}

After the landmark observations of gravitational waves (GWs) from mergers of compact binaries~\cite{GW150914-prl,GW151226-prl,GW170104,GW170608,GW170814,GW170817}, GW astronomy has been established as a brand new way to study the cosmos. The first two observational campaings of Advanced LIGO and Advanced Virgo have provided the first few detections. The current upgrade and commissioning of these detectors will lead to an increase in their sensitivity, in the scale of the cosmic horizon to search for sources, and in the event rates. The upcoming observing run of Advanced LIGO and Advanced Virgo, O3, due to start in early 2019, promises a plethora of new GW data and discoveries \cite{AbbottLR:2018}. 

Despite the recent detections, noise removal remains one of the most challenging problems in GW data analysis. The sensitivity of current ground-based detectors sharply degrades at frequencies below a few tens of Hz due to gravity-gradient (seismic) noise and above $\sim 2$ kHz, due to quantum fluctuations of the laser~\cite{Martynov:2016}. At intermediate frequencies is where intereferometers become the most sensitive, being mainly limited by thermal noise due to Brownian motion of the suspensions and mirrors. Many noise sources affecting detectors are non-Gaussian and non-stationary, altering the sensitivity of the detectors in real time. In addition, transient noise signals of both instrumental and environmental origin, commonly known as `glitches', may not only disturb astrophysical GW signals (as dramatically manifested in the infamous glitch affecting the merger signal of GW170817~\cite{GW170817}) but also mimic true signals, increasing the false-alarm rate and producing a decrease in the detectors' duty cycle. Uninterrupted efforts in detector commissioning and characterization are made to reduce the effects of glitches. In particular, improving the identification and classification of glitches is fundamental to increase the efficiency of the detection. Current efforts employ approaches as diverse as Bayesian inference, Principal Component Analysis, machine learning, deep learning, and even combinations of machine-learning and citizen science (see e.g.~\cite{Powell:2015,Powell:2017,Zevin:2017,Mukund:2017,George:2018,Razzano:2018}). 

In this paper we study the suitability of dictionary-learning algorithms for glitch denoising and classification, taking as starting point the method introduced in our work for GW denoising using dictionaries~\cite{Torres:2016}. Mathematically speaking, a dictionary is a matrix of $m$ \textit{atoms} (signals in our case) of length $n$ organized as columns. The dictionary-learning approach is based on the so-called sparse representation, which states that a given signal can be reconstructed as a linear combination of only a few atoms, $\bm u = \bm D \bm\alpha$, where $\bm u$ is the reconstructed signal, $\bm D = [\bm{d_1}, \dots, \bm{d_m}]$ is the (overcomplete) dictionary composed of $a$ atoms of length $n$ such that $m>n$, and $\bm\alpha\in\mathbb{R}^p$ is a sparse vector containing the coefficients of the representation. Since the dictionary is overcomplete, the solution vector $\bm \alpha$ is not unique, hence we use the \textit{basis pursuit decomposition} proposed in \cite{Chen:2001}. The goal of this paper is to test the suitability of dictionary-learning techniques to denoise glitches and classify them by their morphology. One should note that by ``denoising glitches" we do not mean to remove glitches from the background but to obtain a clear shape from the glitch morphology. In this sense, we treat glitches as signals like we did in our first paper~\cite{Torres:2016}. To test our approach we follow the analysis carried out by~\cite{Powell:2015} where the authors compared three different pipelines to classify simulated glitches embedded in Gaussian noise.

For training the dictionaries and validate our method in a controlled environment, we follow~\cite{Powell:2015} and generate several sets of synthetic glitches which can be classified into three different waveform morphologies: sine Gaussian (SG), Gaussian (G) and Ring-Down (RD). All glitches to be denoised and classified are injected into simulated non-white Gaussian noise similar to that of advanced LIGO in the proposed broadband configuration with different values of the signal-to-noise ratio (SNR).

The classification method we introduce in this paper uses dictionaries in two steps. First, we perform the denoising step aiming for the most faithful reconstructions using three trained dictionaries, one dictionary per waveform morphology. Due to our signal configurations (which will be justified later on) all three dictionaries are likely to yield reconstructions from all glitches, regardless of their morphology. For example, a SG glitch (injected in noise) may be reconstructed by all three dictionaries with relative fidelity. However, the reconstruction from a SG dictionary should be closer to a clean model of a SG glitch than those produced by both the G and RD dictionaries, whose atoms contain glitches of other waveform morphologies. Motivated by this hypothesis, in the second step of our procedure we look for the dictionary whose reconstruction is closer to a perfect waveform morphology, for each unknown glitch to be classified  (i.e.~closer to an ideal SG, G, or RD glitch). In order to achieve this we use three untrained dictionaries whose atoms are whole centered glitches, each one serving as examples of ideal glitches of a single waveform morphology. Every denoised glitch from the first step is reconstructed by all three untrained dictionaries using as few atoms as possible, so that instead of generating a faithful reconstruction each dictionary will yield a ``reconstruction'' close to an ideal glitch of its own morphology while trying to get as close to the denoised glitch as possible. Finally, the predicted waveform morphology of an unknown glitch will be that of the untrained dictionary that produced the reconstructions closest to the three respective reconstructions of the denoising dictionaries. Our study shows that learned dictionaries are an interesting new approach for glitch classification as we manage to successfully classify about 96\% of simulated glitches for a large range of initial parameters. 

This paper is organized as follows. In section \ref{section:math} we summarize the mathematical background of the sparse representation and dictionary learning techniques. In section \ref{sec:signals} we describe the morphology of the three classes of simulated glitches and the parameters that define them. The classification algorithm we have developed using the dictionaries is presented in section \ref{sec:classification}. In section \ref{sec:test} we describe the tests we have performed in order to determine the best set of parameters of the method that produces optimal results and we discuss the application of these techniques (denoising + classification) to a long run of 3000 glitches. Finally, in section \ref{sec:conclusions} we present the main conclusions of our work and outline possible future directions of research. 

\section{Mathematical framework}
\label{section:math}

\subsection{Sparse reconstruction}

In~\cite{Torres:2016} we considered sparse reconstructions of GW signals over trained dictionaries, employing numerical relativity catalogs of core-collapse supernova signals and binary black hole waveforms. In this paper, we apply the same approach to classify and reconstruct simulated GW glitches embedded in Gaussian noise. Similar to our work on GW denoising using dictionaries~\cite{Torres:2016} we assume that the way in which glitches are embedded into noise can be described by the linear degradation model
\begin{equation}
\bm{f} = \bm{u} + \bm{n}\,,
\end{equation}
where $\bm{f} \in \mathbb{R}^n$ is the data from the detector, $\bm{u} \in \mathbb{R}^n$ is the glitch to be recovered for later classification, and $\bm{n} \in \mathbb{R}^n$ is random Gaussian noise. Given an overcomplete dictionary $\bm{D} \in \mathbb{R}^{n\times m}$, where the number of atoms $m$ is greater than their length $n$, there is a sparse vector $\bm{\alpha} \in \mathbb{R}^m$ for which $\bm{D}\bm{\alpha} \sim \bm{u}$, where 
\begin{equation} \label{eq:unconstrained}
\bm{\alpha} = \argmin_{\bm{\alpha}} \left\{ 
  \Vert \bm{f} - \bm{D\alpha} \Vert^2_2
  + \lambda \Vert\bm{\alpha}\Vert_0
\right\}\,.
\end{equation}
In this equation $\Vert \cdot \Vert_0$ and $\Vert \cdot \Vert_2$ stand for the $L^0$-norm and the $L^2$-norm, respectively. The former is just the number of nonzero components of its argument. 
This problem can be written as a convex and unconstrained variational problem by substituting the $L^0$-norm of $\bm{\alpha}$ by the $L^1$-norm as a  penalty term of the problem weighted by a Lagrangian multiplier $\lambda$,
\begin{equation} \label{eq:lasso}
\bm{\alpha} = \argmin_{\bm{\alpha}} \left\{ 
  \Vert \bm{f} - \bm{D\alpha} \Vert^2_2
  + \lambda \Vert\bm{\alpha}\Vert_1
\right\}\,,
\end{equation}
an approach which is known as \textit{basis pursuit}~\cite{Chen:2001} or \textit{LASSO}~\cite{lasso}. The regularization in the $L^1$-norm
promotes zeros in the components of the vector coefficient $\bm{\alpha}$ and, thus, the solution of this variational problem is typically the sparsest one.

The Lagrangian multiplier $\lambda$ is also called the regularization parameter, as it regulates the level of detail to be recovered in the sparse representation $\bm\alpha$ of the input signal $\bm{f}$. The higher the value of $\lambda$ the more the $L^1$-norm term weights, making the coefficients  of $\bm\alpha$ tend to zero when solving problem \eqref{eq:lasso}, which results in less atoms being used for the sparse representation (i.e.~for the reconstruction of $\bm{f}$). On the other hand, using low values of $\lambda$ favours the {\it fidelity} term (the $L^2$-norm term in Eq.~(\ref{eq:lasso})), which results in more (or even all) atoms being used. This  transforms the problem into a simple least-squares problem and yields a reconstruction as similar as possible to the input data. The optimal value for a given signal, $\lambda_{\text{opt}}$, is defined to be the one which gives the bests results according to a suitable metric function applied to the denoised signal and to the original one, measuring the quality of the recovered signal. In this work we use two estimators, namely the Mean Squared Error,
\begin{equation}
\text{MSE} = \frac{1}{n} \sum_{i=1}^n (\hat{Y}_i - Y_i)^2~,
\end{equation}
where $\hat{Y}$ and $Y$ are the reconstructed and original signals respectively, and $n$ is the number of samples, and the structural similarity (SSIM) index~\cite{Wang:2004} which takes into account the structural information. The SSIM index varies between -1 (minimum similarity) and 1 (maximum similarity) and is defined as
\begin{equation}
\text{SSIM}(x,y) = \frac{(2\mu_x\mu_y + c_1)(2\sigma_{xy} + c_2)}{(\mu_x^2 + \mu_y^2 + c_1)(\sigma_x^2 + \sigma_y^2 + c_2)} ~,
\end{equation}
where $c_1$ and $c_2$ are constants, $\mu_x$ ($\mu_y$) is the average of $x$ ($y$), $\sigma_x^2$ ($\sigma_y^2$) the variance of $\mu_x$ ($\mu_y$) and $\sigma_{xy}$ the covariance of $x$ and $y$. In order to facilitate the comparison between the two estimators we rewrite the SSIM index so that its value ranges from 0 (maximum similarity) to 1 (minimum similarity), like the MSE,
\begin{equation}
\text{DSSIM}(x,y) = \frac{1 - \text{SSIM}(x,y)}{2}~.
\end{equation}
This metric function is called the structural dissimilarity index.

To solve the LASSO problem we use the modified least-angle regression (LARS) algorithm~\cite{Efron:2004}. This algorithm is similar to a forward stepwise regression: it starts with the regression coefficients equal to zero, and for each iteration it finds the predictor $u_i$ most correlated with the response $f$ and takes a large step in the same direction until other predictor achieves a similar correlation. However, instead of continuing along the first predictor, LARS proceeds following recursively the equiangular direction between the predictors of equal correlation. The main advantage of this method is the efficiency when dealing with more dimensions than points ($p > n$), which is the case of overcomplete dictionaries (see~\cite{Efron:2004} for further details). 

\subsection{Dictionary learning problem}

Before the denoising, the dictionaries are trained with a set of signals splitted in $p$ patches of length $n$, $\bm{U} = [u_1, \dots, u_p] \in \mathbb{R}^{n\times p}$. The number of training patches is large compared with the number of atoms and their length, $p \gg m,n$, because of the sparsity condition and the overcompleteness of the dictionary.

The trained dictionary is obtained by adding the dictionary matrix $\bm{D}$ as a variable in the minimization problem,
\begin{equation}
\bm{\alpha} = \argmin_{\bm{\alpha},\bm{D}} \frac{1}{n} \sum_{i=1}^p \left\{
  \Vert \bm{u}_i - \bm{D}\bm{\alpha_i} \Vert_2^2 + \lambda\Vert\bm{\alpha}_i\Vert_1
\right\} ~,
\end{equation}
where the summation index $i$ indicates the \textit{i}-th row of matrix $\bm{\alpha}\in\mathbb{R}^{p\times n}$, which contains the coefficients of the sparse representation of each atom. The columns $(\bm{d}_i)_{i=1}^p$ of the dictionary are constrained to have an $L^2$-norm less or equal to one, $\bm{d}_i^T \bm{d}_i \leq 1$, to prevent $\bm{D}$ from being arbitrarly large.

The problem is solved by the algorithm proposed by Mairal et al in~\cite{Mairal2009} with the mini-batch optimization. This is a block-coordinate descend method which minimizes $\bm{D}$ and $\bm{\alpha}_i$ separately for each iteration $t$,
\begin{align}
\bm{\alpha}^t =& \argmin_{\bm{\alpha}} \left\{
  \frac{1}{2} \Vert \bm{u}_t - \bm{D}^{t-1}\bm{\alpha} \Vert_2^2
  + \lambda \Vert\bm{\alpha}\Vert_1
\right\}\,,
\\
\bm{D}^t =& \argmin_{\bm{D}} \frac{1}{t} \sum_{i=1}^t \left\{
  \frac{1}{2} \Vert \bm{D}\bm{\alpha}_i^t - \bm{u}_i \Vert_2^2
  + \lambda \Vert\bm{\alpha}_i\Vert_1
\right\} ~,
\end{align}
 with the advantage of being parameter-free and not requiring any learning rate.

\begin{figure}[t]
\centering
\includegraphics[width=10cm]{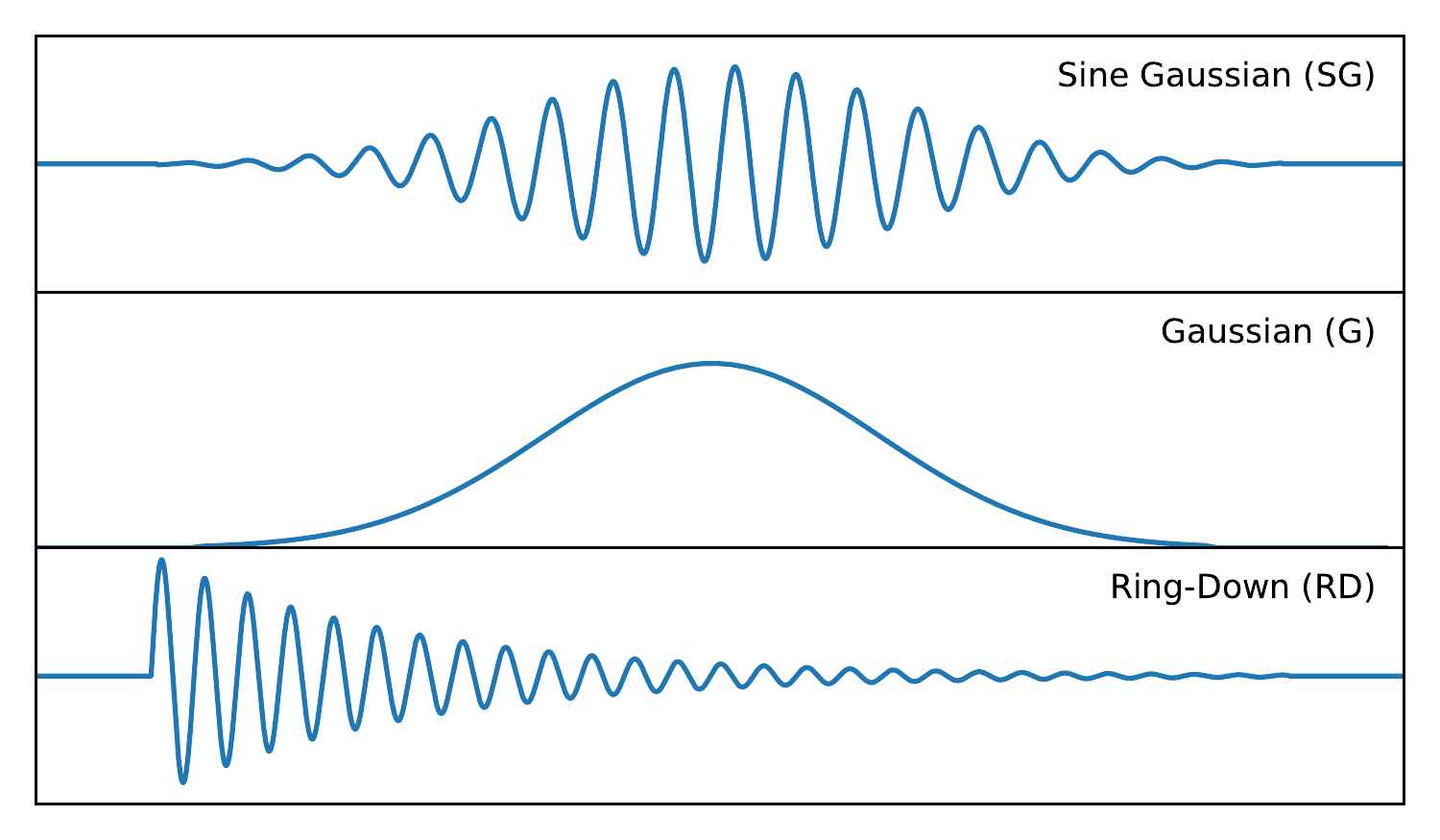}
\caption{\label{fig:morphology} Examples of the three different glitch morphologies used in our data sets.}
\end{figure}

\section{Signal set}
\label{sec:signals}

In this work we do not use actual detector glitches. Instead, following~\cite{Powell:2015}, we simulate three simple kinds of glitch morphologies, namely sine Gaussian (SG), Gaussian (G) and ring-down (RD). They are defined by the following functions
\begin{align} \label{eq:glitch_equations}
h_{\rm SG}(t) =& ~h_0 \sin{\left\{ 2\pi f_0(t-t_0) \right\}} e^{-(t-t_0)^2/2\tau^2}~, \\
h_{\rm G}(t) =& ~h_0 e^{-(t-t_0)^2/2\tau^2}~, \\
h_{\rm RD}(t) =& ~h_0 \sin\{ 2\pi f_0(t-t_0) \} e^{-(t-t_0)/2\tau}~,
\end{align}
where $f_0$ is the central frequency, $t_0$ is a characteristic time for each of the waveforms, namely the time at the centre of the SG and G waveforms and at the beginning of the RD waveform, $\tau = Q/\sqrt{2}\pi f_0$, with $Q$ being the quality factor, and 
$h_0 = {h_{\rm rss}}/\sqrt{\tau}$, with $h_{\rm rss}$ being the root sum squared amplitude of the glitch.
These parameters are randomly chosen within the ranges shown in Table~\ref{tab:glitch_params}, with a linear distribution on their logarithms to get enough samples of all orders of magnitude.
\begin{table}
\caption{\label{tab:glitch_params}Minimum and maximum parameters of the simulated glitches, from~\cite{Powell:2015}.}
\begin{indented}
\item[]\begin{tabular}{@{}llll}
\br
&Waveform&Minimum&Maximum\\
\mr
$f_0$ (Hz)&All&40&1500\\
$h_{\rm rss}$ (Hz$^{-1/2}$)&All&$5\times10^{-22}$&$4\times10^{-21}$\\
$Q$ &SG, RD&2&20\\
Duration (s)&G&0.001&0.01\\
\br
\end{tabular}
\end{indented}
\end{table}
Examples of the waveform morphology for all three glitches are shown in Figure~\ref{fig:morphology}. The SG and RD waveforms are relatively similar between them, both displaying a distinctive oscillatory pattern, and more complex than the G waveform. This variety of signals serves the purpose of testing the capability of the classification algorithm in a realistic scenario involving different types of glitches.

We generate three separate sets of glitches with the sampling rate of the Advanced LIGO/Virgo detectors, 16384 Hz. Since for training each dictionary we want to use the same amount of training patches (namely 20000), and each morphology has its own range of durations (which translates to different range of samples), we generate different amounts of glitches. Moreover, we will test dictionaries with different atoms' lengths. Therefore, we also need to be sure we will be able to generate enough patches with the largest atoms. Taking both conditions into account, we generate a training set of 915 SG glitches, 62440 G glitches, and 603 RD glitches. These signals are normalized to their $L^2-$norm to ensure the best convergence conditions for the learning algorithm. Then we use a set of 300 glitches (100 per waveform morphology) for parameter optimization, and a bigger set of 3000 glitches (1000 glitches per waveform morphology) to test the final classification algorithm. All signals from these two sets are scaled so that their maximum value is equal to one.

The glitches used for training and testing are injected into white Gaussian noise weighted by the power spectral density (PSD) of Advanced LIGO in the proposed broadband configuration, as explained in~\cite{Torres:2014}. We rescale the signals to a SNR value of 20. The SNR is defined as
\begin{equation}
\text{SNR} = \sqrt{4 \Delta t^2 \Delta f \sum_{k=1}^{N_f} \frac{\vert \tilde{u}(f_k) \vert^2}{S(f_k)}} ~,
\end{equation}
where $\tilde{u}$ indicates the Fourier transform of signal $u$, $S$ is the PSD of the noise corresponding to the sensitivity curve of the detector, $f_k$ is each of the components of the frequency vector, $N_f$ is the number of positive frequencies, and $\Delta t$ and $\Delta f$ are the time step and frequency step, respectively.

\section{Classification method}
\label{sec:classification}

\begin{figure}[t]
\centering
\includegraphics[scale=1.0]{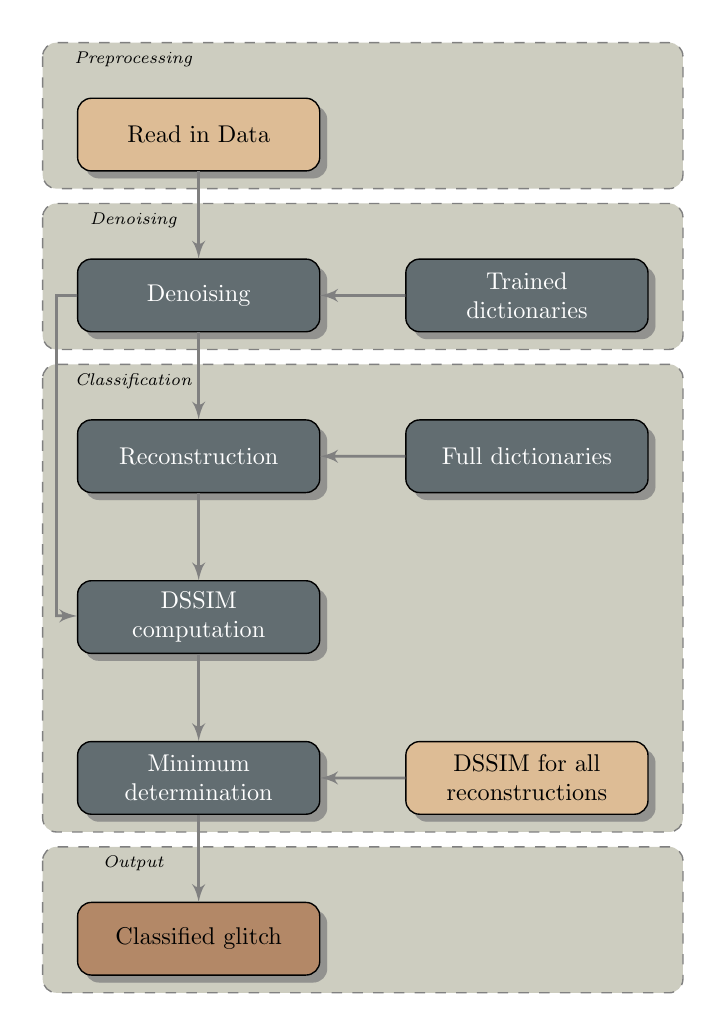}
\caption{\label{fig:classification_diagram} Block diagram outlining the classification method.}
\end{figure}

The classification method we propose in this work makes use exclusively of dictionaries for sparse coding, and can be divided in two phases: the denoising phase and the discrimination phase. For convenience, a block diagram outlining the classification method that we discuss next is displayed in Figure~\ref{fig:classification_diagram}.

In the first phase the goal is to recover as much oscillations as possible from all glitches while keeping the spurious oscillations to a reasonable low level (let us call them `parent' reconstructions). To this end we use three trained dictionaries, one per waveform morphology, with a constant value for the regularization parameter $\lambda_{\text{tr}}$ (the subindex `tr' stands for `transformation' as it refers to the sparse encoding of the signals). Each denoising dictionary is composed of atoms of the same length, which are shorter than the length of the input signals. Because of this, part of the atoms are fragments of glitches, and none of them are aligned in any way. Therefore, each dictionary is initialized and trained with a large number of patches randomly extracted from the corresponding training set, with the only condition that the patches include a minimum number of nonzero bins, which we choose to be around \sfrac{1}{4} of the atom's length. Since we do not know the glitch morphology yet, each glitch needs to be reconstructed by all denoising dictionaries.

After this phase we end up with three denoising reconstructions per unknown glitch. Ideally, each dictionary should be used with an optimal regularization parameter $\lambda_{\text{opt}}$ so that the dictionary would yield a relatively clean reconstruction only if the input signal contained a glitch of the same morphology. However, in practice some factors like a high resemblance between some morphologies or waveforms with extremely low SNR make impossible for our dictionaries to achieve an acceptable level of glitch discrimination. To overcome this issue we take a different approach with respect to our previous work \cite{Torres:2016}; we redefine the value of $\lambda_{\text{opt}}$ to be that which gives us the best classification results with our set of testing glitches. As shown in section \ref{subsec:parameter_optimization}, this new value happens to be quite lower than the original $\lambda_{\text{opt}}$, making all dictionaries to yield relatively faithful reconstructions regardless of the morphology of the glitch. However, given a noisy signal containing (for example) a SG glitch, we expect the reconstruction from the SG dictionary to be closer to a perfect SG waveform than those from the G and RD dictionaries, since the first one is made up of patches of perfect glitches of the same morphology. In other words, for each glitch we need to determine which denoising dictionary produces the reconstruction closest to an ideal glitch of any of the three morphologies.

\begin{figure}[!ht]
\centering
\includegraphics[scale=1.4]{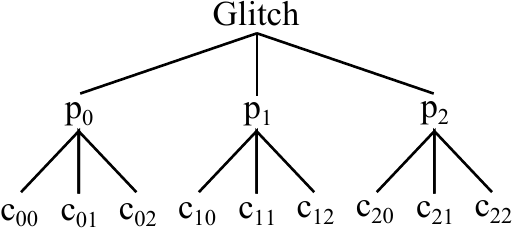}
\caption{\label{fig:classification_tree} Tree diagram of the reconstructions obtained for each original glitch, where $p_j$ stands for parent glitches, and $c_{ji}$ for children glitches. The numbering corresponds to the three morphologies ordered as follows: sine Gaussian ($i,j = 0$), Gaussian ($i,j = 1$), Ring-Down ($i,j = 2$).}
\end{figure}

The goal of the second phase is to perform the aforementioned discrimination using untrained dictionaries, which we call classification dictionaries, composed of atoms containing whole centered glitches. Each of these dictionaries is nothing more than a collection of random examples of ``perfect'' glitches belonging to the same waveform morphology. With them, every parent will be ``reconstructed'' using as few atoms as possible, so that instead of generating a faithful reconstruction, each dictionary will choose the few atoms that most ``resemble'' to the parent glitch but maintaining the morphology of the dictionary. The number of atoms used, which will be referred to as `nonzeros' from now on, is a hyperparameter which will be fixed instead of $\lambda_{\text{opt}}$, and whose optimum value needs to be determined. After this procedure, we are left with three new reconstructions (let us call this new reconstructions ``children'') per parent glitch, with a total of 12 reconstructions for each unknown glitch, forming a ternary tree structure as shown in figure~\ref{fig:classification_tree}. 

Finally, we use the SSIM estimator to calculate the value of the DSSIM index between each child and its parent. In this part of the algorithm we do not make use of the MSE estimator because it does not take into account the structural information. The predicted morphology will be that of the dictionary which produced the least total DSSIM value,
\begin{equation}
i_{\text{dict}} = \argmin_i \prod_j \text{DSSIM}(p_j, c_{ji})~,
\end{equation}
where $i_{\text{dict}}$ is the index of the dictionary. If a child $c_{ji}$ is not reconstructed (i.e.~it is not recognized) the DSSIM will take the worst value, and if none of the children or parents are reconstructed, then we will consider the glitch lost (which means it will not be classified whatsoever).

\section{Tests and results}
\label{sec:test}
\subsection{Parameter optimization} \label{subsec:parameter_optimization}

We first optimize the parameters of the denoising dictionaries, namely the regularization parameter of the learning step $\lambda_{\text{learn}}$, the number of atoms $m$ and their lengths $n$, and the regularization parameter of the reconstruction step $\lambda_{\text{tr}}$. We then next optimize the parameters of the classification dictionaries, namely the number of atoms and the number of nonzero coefficients to use for the reconstructions of the second phase. For each parameter we test several different values, reconstructing all 100 glitches from the validation set and computing the average MSE (similar results are obtained when using the DSSIM index, since now we are not comparing between different morphologies). We consider the optimum value of a parameter to be the one that yields the lowest average MSE. 

\begin{figure}[!ht]
\centering
\includegraphics[width=8cm]{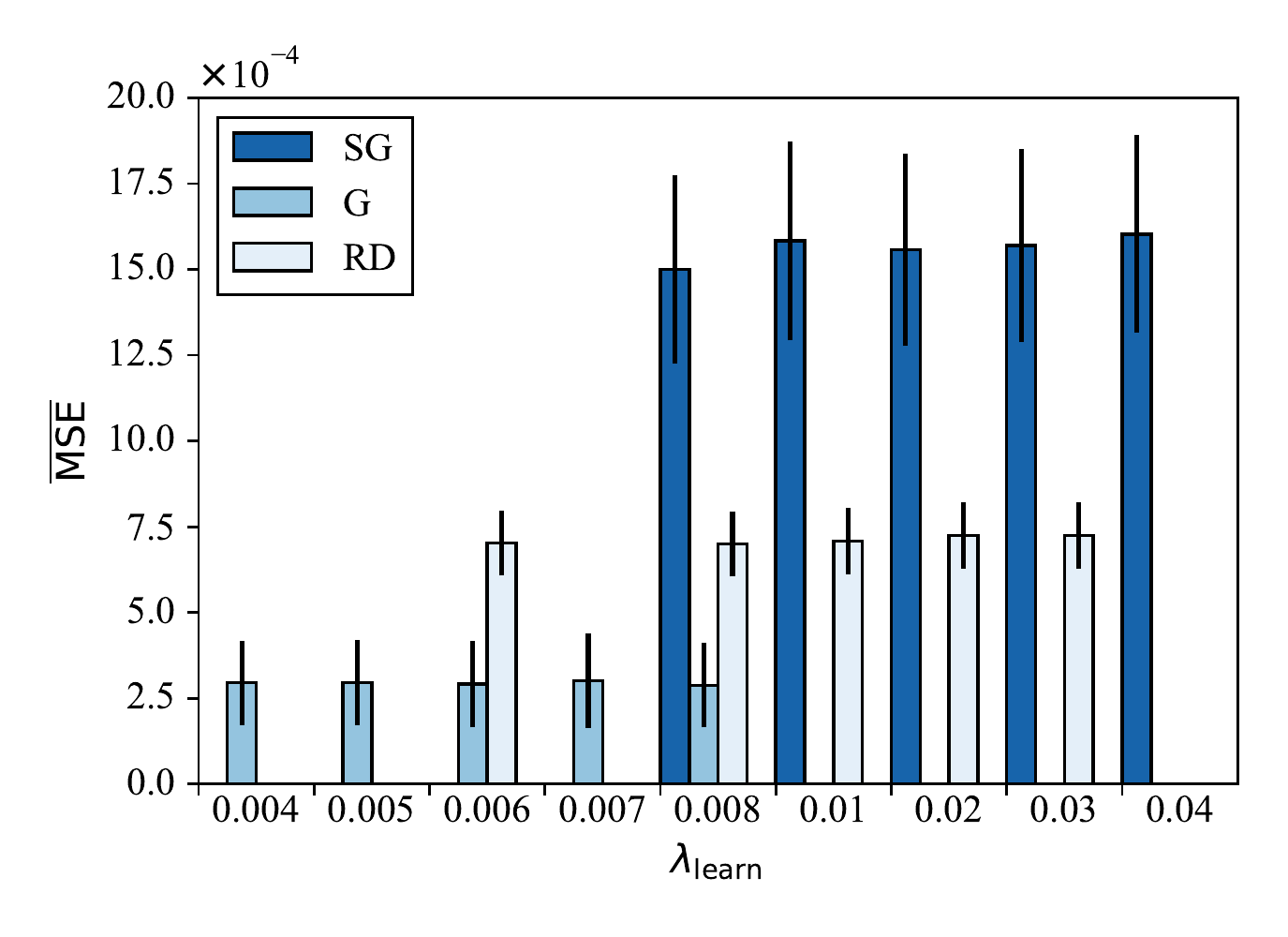}
\caption{\label{fig:lambda_learn} Histogram of the average MSE for all validation signals reconstructed by the denoising dictionaries trained with different values of $\lambda_{\text{learn}}$. Each colour represents a different dictionary (i.e.~a different glitch morphology). The standard deviation of each average is represented by a black line, and downscaled by a factor 10 for SG and RD.}
\end{figure}

We start by studying how the regularization parameter of the dictionary learning step, $\lambda_{\text{learn}}$, affects the reconstructed signals. As a starting point we choose 1024 atoms for the SG and RD dictionaries with a length of 512 samples, and 256 atoms for the G dictionary with a length of 128 samples. Those lengths are close to the validation set's average. An initial test shows  that we obtain trained atoms without noise only when $\lambda_{\text{learn}}$ is between a certain range, which is different for each morphology as shown in figure~\ref{fig:lambda_learn}. The absence of a histogram for certain values of $\lambda_{\text{learn}}$ indicates that the dictionary does not yield a reconstruction for that value.
However, there is no significant variation in the quality of the reconstruction for different values of $\lambda_{\text{learn}}$ (once a reconstruction has been obtained). This can be seen by comparing the histogram values with their standard deviation. Therefore, we choose to use a mean value for each dictionary: 0.02 for SG, 0.006 for G, and 0.01 for RD.

\begin{figure}[!ht]
\centering
\includegraphics[width=8cm]{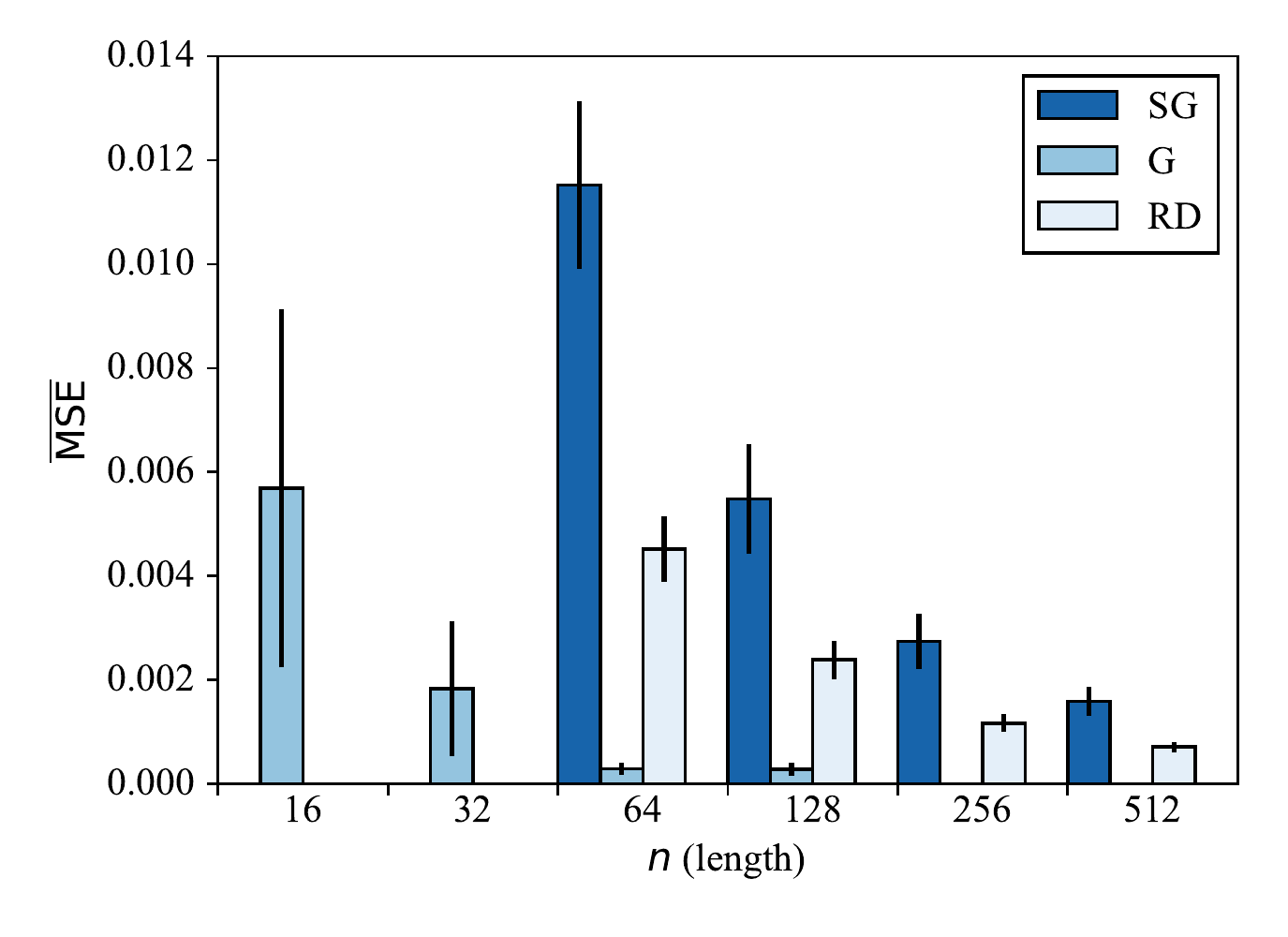}
\caption{\label{fig:length_atoms} Histogram of the average MSE for all validation signals reconstructed by the denoising dictionaries with different atoms' lengths. Each colour represents a different dictionary (i.e.~a different morphology). The standard deviation of each average is represented by a black line, and downscaled by a factor 10 for SG and RD. }
\end{figure}

Next, we test the effect of using different atom lengths, $n =$ 64, 128, 256 and 512 for SG and RD dictionaries, and $n =$ 16, 32, 54 and 128 for G dictionary (Gaussian glitches are shorter than the other two). The results displayed in figure \ref{fig:length_atoms} show that shorter atoms produce worse reconstructions (i.e.~larger values of the MSE index). This is due to the fact that if atoms are too short they are more sensible to noise oscillations, becoming hard to recognize low frequency glitches. The best results are achieved with the two greatest lengths of each dictionary. Therefore, we choose to use a length of 256 for SG and RD dictionaries, and 128 for the G dictionary, which gives a good trade-off between quality and performance.

\begin{figure}[!ht]
\centering
\includegraphics[width=8cm]{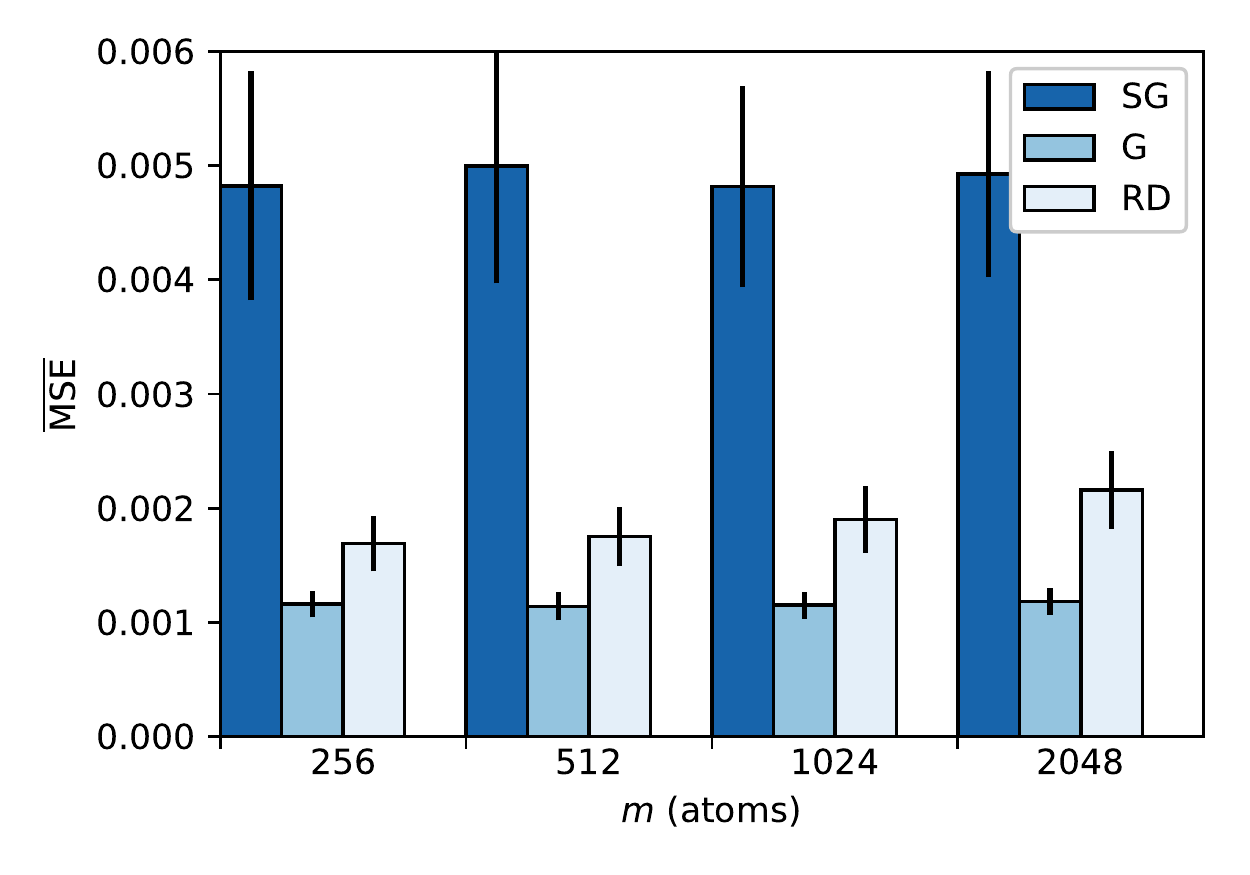}
\caption{\label{fig:number_atoms} Histogram of the average MSE for all validation signals reconstructed by the denoising dictionaries with different number of atoms. Each colour represents a different dictionary (i.e. a different morphology). The standard deviation of each average is represented by a black line, all downscaled by a factor 10.}
\end{figure}

In general, the larger the number of atoms the better the results should be. To check this, we carry out a test with $m=$ 256, 512, 1024 and 2048 atoms. We set all their lengths to $n = 128$ (only for this test) so that dictionaries remain over-completed. As can be seen in figure~\ref{fig:number_atoms} there is no clear improvement; even 256 atoms are more than enough for the dictionaries to recognize their own waveforms, which explains why increasing their number does not provide better results anymore. Therefore, we use the smallest number of atoms needed for the dictionaries to remain overcomplete: $m =$ 512 for SG and RD, and 256 for Gaussian.

\begin{figure}[!ht]
  \centering
  \includegraphics[width=0.49\textwidth]{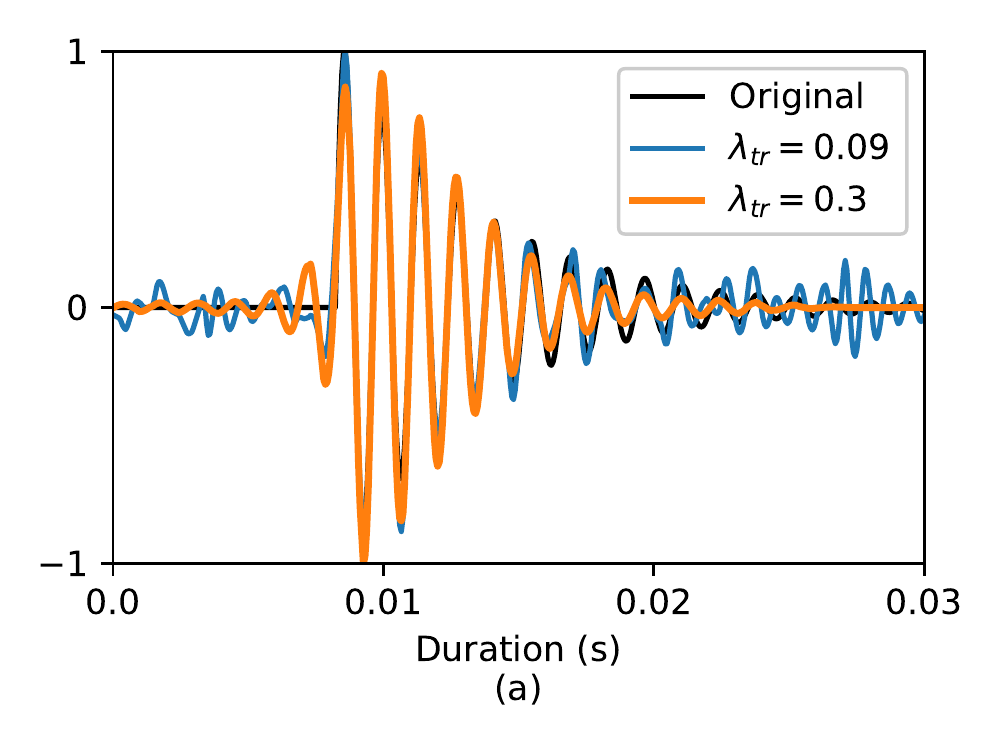}
  \includegraphics[width=0.49\textwidth]{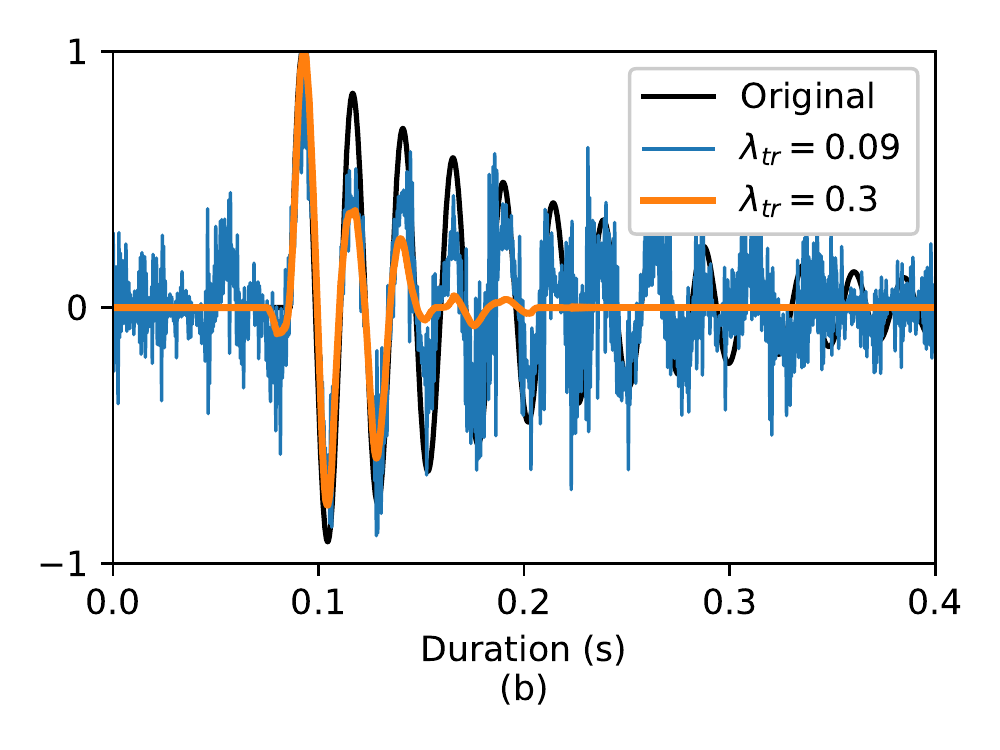}
  \caption{\label{fig:examples_denoising} Example of two denoised RD glitches which were injected into Gaussian noise at 20 SNR. Both plots show the original glitch (before being injected into noise), and two reconstructions using different values for $\lambda_{\text{tr}}$, namely 0.3, which is close to the average optimal value and yields almost noise-free reconstructions, and 0.09, which improves the classification properties of the dictionaries but introduces more spurious oscillations.}
\end{figure}

Every glitch signal to be denoised has a specific optimum value of the regularization parameter, $\lambda_{\text{opt}}$, used in its sparse reconstruction (see~\cite{Torres:2016} for details on the computation of $\lambda_{\text{opt}}$). However, our initial tests showed that although using a mean value of $\lambda_{\text{opt}}$ (estimated from the validation set) yielded the most accurate reconstructions, it did not offer the best classification results. This is because the distinctive oscillatory features of a large number of glitches of our sample cannot be fully recovered with the optimum values of the regularization parameter, as e.g.~happens with the RD glitch shown in figure \ref{fig:examples_denoising}(b). Instead, more oscillations can be recovered by using a smaller value, $\lambda_{\text{tr}}$, making the glitch morphology more easy to differentiate (and, hence, improving glitch classification) but at the expense of a poorer denoising. Nevertheless, those glitches that could be completely reconstructed with the optimum value (like the example shown in figure \ref{fig:examples_denoising}(a)) can still be partly reconstructed using $\lambda_{\text{tr}}$. Therefore, we carry out a test to find a new (average) optimum value for the regularization parameter of the transformation. We use the same value for all three dictionaries in order to spend a reasonable amount of computational time. More precise classifications may be achieved by looking for individual values, although judging by our tests we would not expect a big improvement.

\begin{figure}[!ht]
  \centering
  \includegraphics[width=5.1cm]{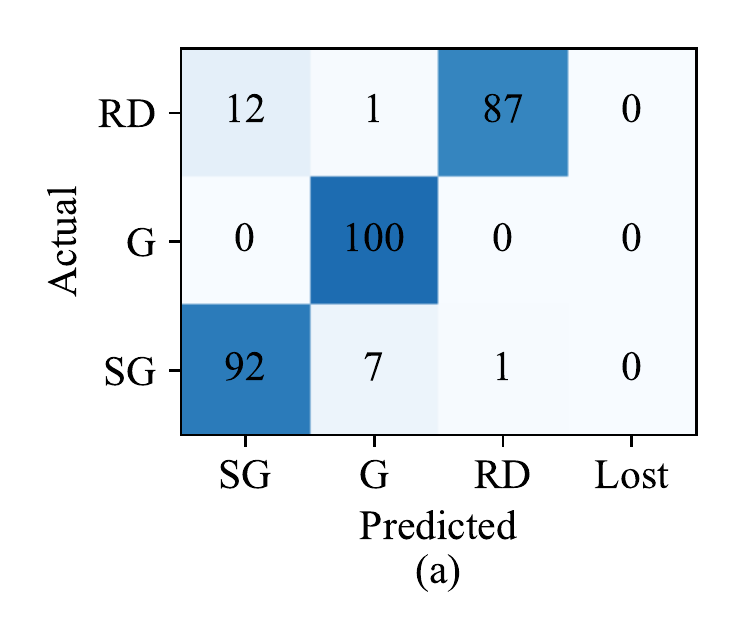}
  \includegraphics[width=5.1cm]{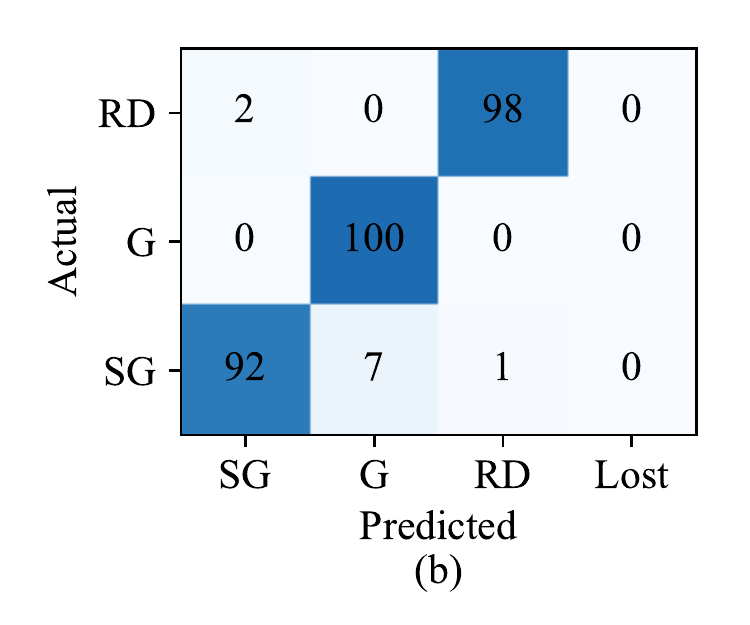}
  \includegraphics[width=5.1cm]{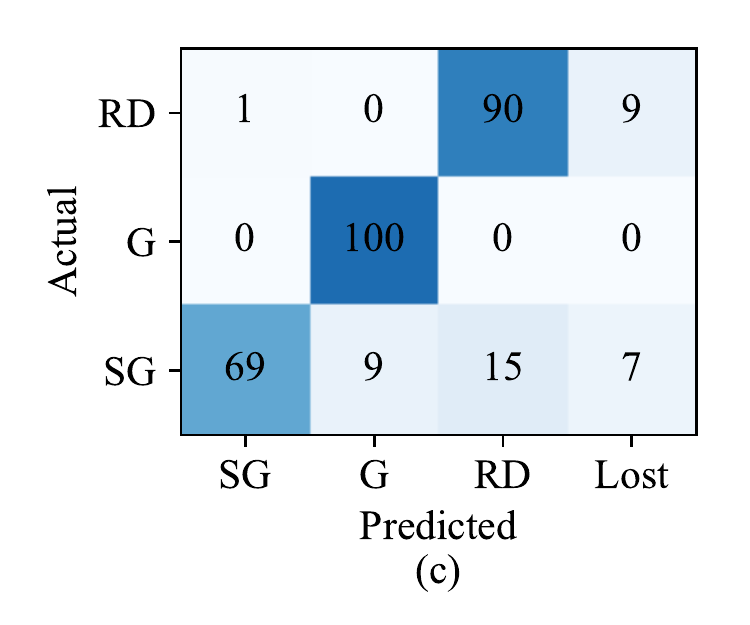}
  \caption{\label{fig:lambda_transform} Confusion matrices of the classification results for three different values of the regularization parameter of the transformation of the denoising dictionaries: $ \lambda_{\text{tr}}$ =  0.05 (a), 0.09 (b), 0.5 (c). Rows correspond to the actual morphology of validation glitches, and columns to the morphology predicted by our classification dictionaries. We chose as a starting values 256 atoms for the classification dictionaries, and 4 nonzero coefficients for their reconstructions.}
\end{figure}
\begin{figure}[!ht]
  \centering
  \includegraphics[width=5.5cm]{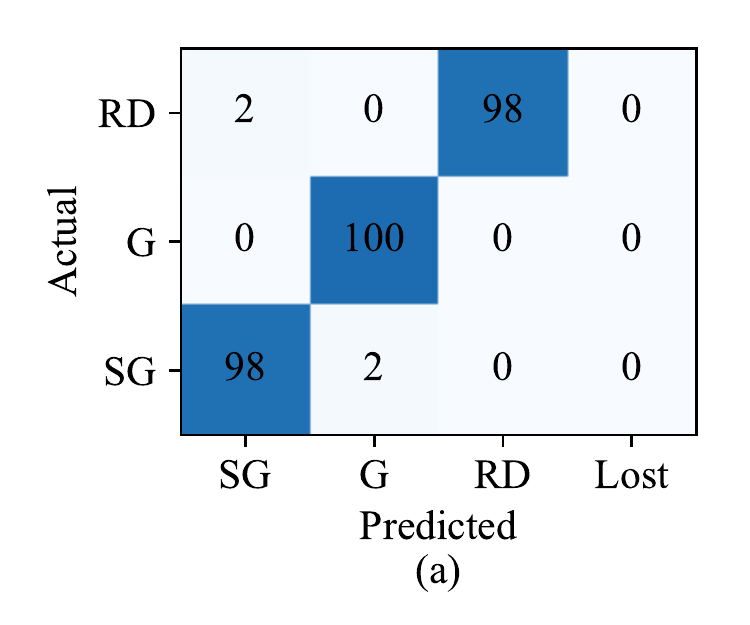}
  \includegraphics[width=5.5cm]{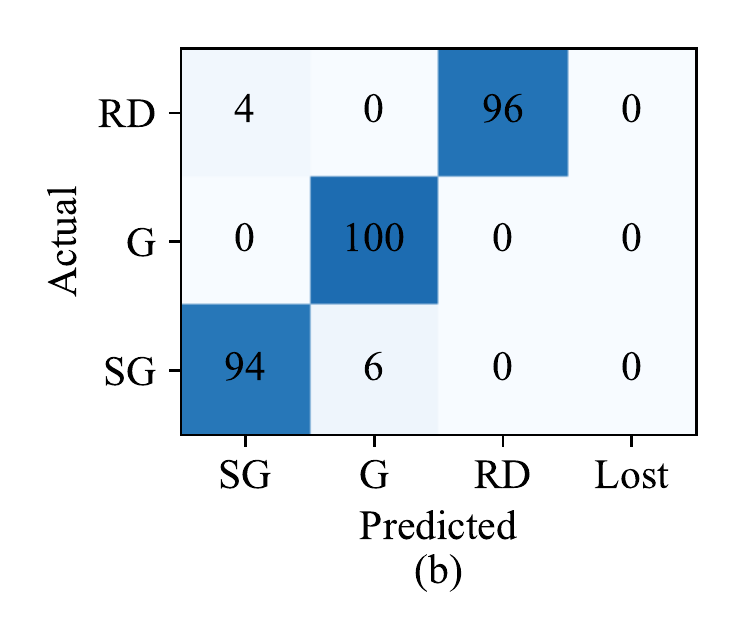} \\
  \includegraphics[width=5.5cm]{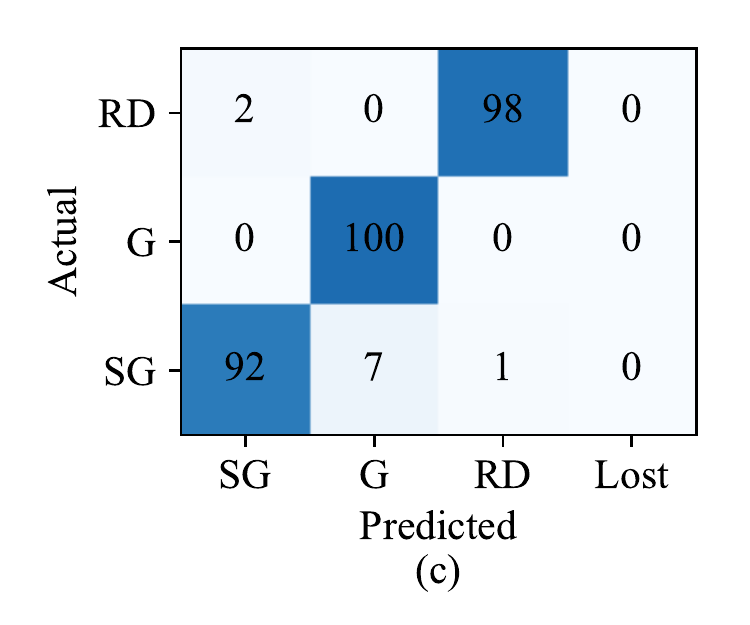}
  \includegraphics[width=5.5cm]{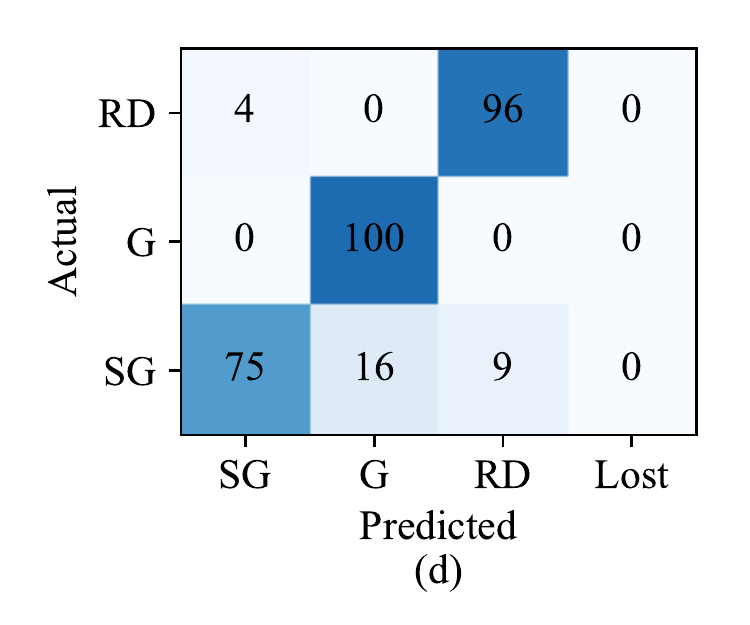}
  \caption{\label{fig:nonzeros} Confusion matrices of the classification results using different numbers of nonzero coefficients: 1 (a), 2 (b), 4 (c), and 8 (d). Rows correspond to the actual morphology of validation glitches, and columns to the morphology predicted by classification dictionaries. We keep using 256 atoms as a starting value for the classification dictionaries.}
\end{figure}
 From all tested values, only two extremes and the optimum result are shown in the confusion matrices of figure \ref{fig:lambda_transform}. All Gaussian glitches are always correctly classified due to their simplicity compared to SG and RD glitches. The latter two are sometimes mismatched because of their similarity. When the value of $\lambda_{\text{tr}}$ of the denoising dictionaries is too low (figure \ref{fig:lambda_transform}a) more RD glitches are predicted as SG glitches, which means that the SG dictionary is more capable of reproducing the RD morphology than otherwise. On the other hand, when $\lambda_{\text{tr}}$ is too high (figure \ref{fig:lambda_transform}c) some glitches are lost (because the reconstruction capability of the dictionaries is reduced), and more SG glitches are mistaken as RD glitches because the number of irregular reconstruction increases (improving the chances of the RD dictionary to provide better reconstructions of the denoised glitches). The best configuration found is $\lambda_{\text{opt}} = 0.09$ (figure \ref{fig:lambda_transform}b) for all denoising dictionaries, with no glitches lost and only a few and fairly well balanced mismatches.

For the classification dictionaries we need to choose the number of nonzero components we will be using for the second reconstruction step. The goal of these dictionaries is to discriminate between the different morphologies by reconstructing again the denoised glitches and finding the best fit. Hence we are not interested in using too many atoms in each reconstruction (which eventually would lead to too similar reconstructions from different dictionaries). In fact, the results plotted  in figure \ref{fig:nonzeros} show that the best discrimination level is achieved with only 1 nonzero coefficient (figure \ref{fig:nonzeros}a), while the worst results are obtained with 8 nonzero coefficients (figure \ref{fig:nonzeros}d) as expected.

\begin{figure}[!ht]
  \centering
  \includegraphics[width=5.5cm]{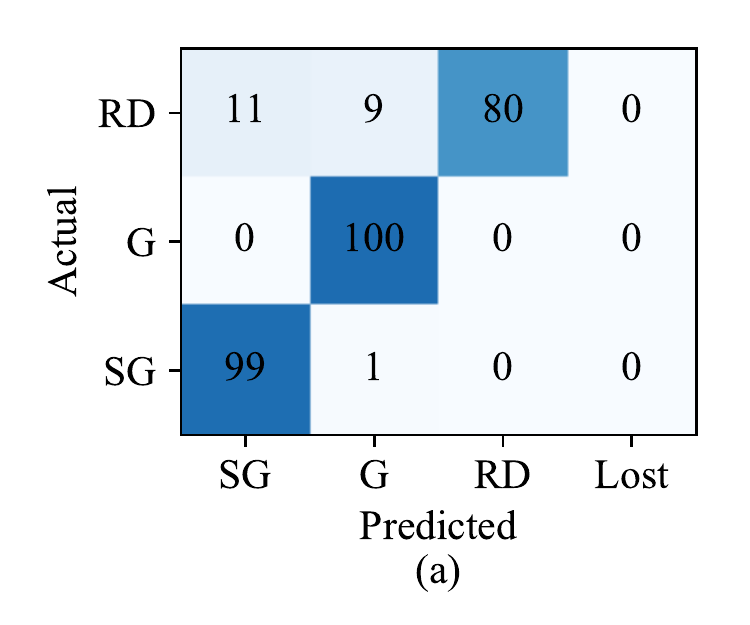}
  \includegraphics[width=5.5cm]{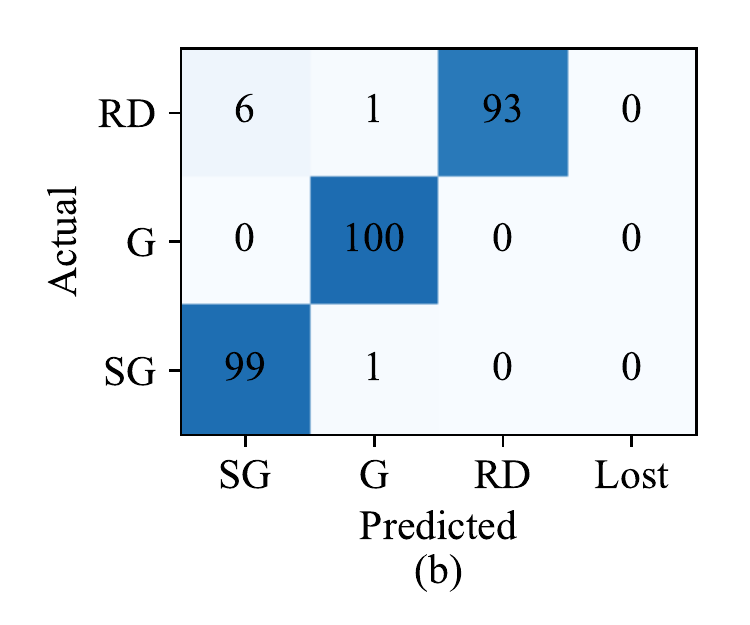} \\
  \includegraphics[width=5.5cm]{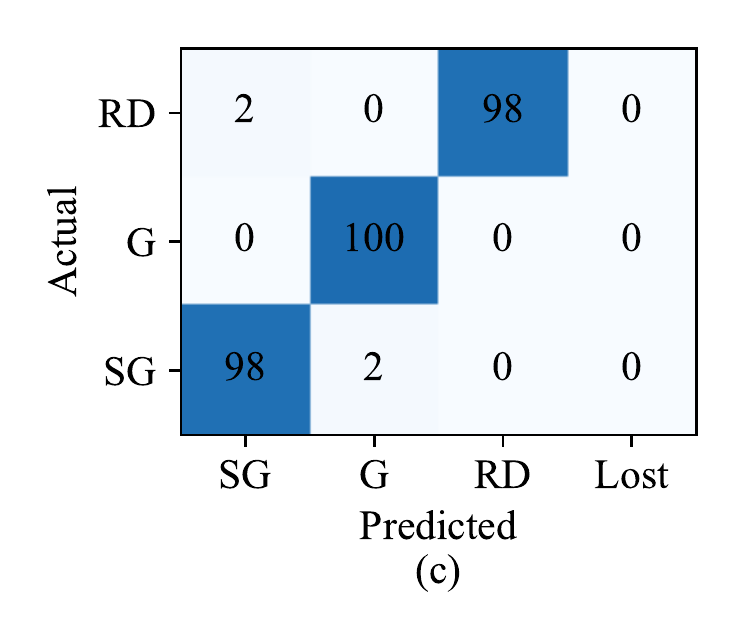}
  \includegraphics[width=5.5cm]{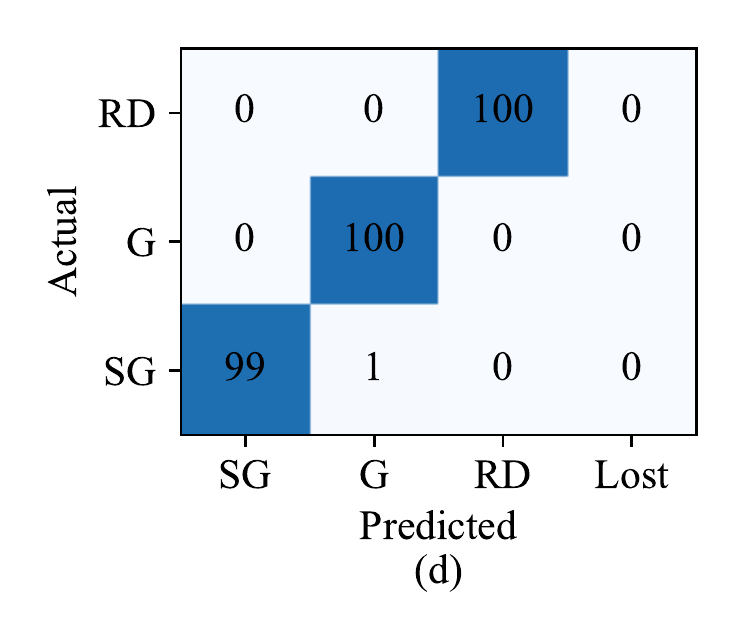}
  \caption{\label{fig:natoms} Confusion matrices of the classification results using different numbers of atoms: 64 (a), 128 (b), 256 (c), and 512 (d). Rows correspond to the actual morphology of validation glitches, and columns to the morphology predicted by classification dictionaries.}
\end{figure}
Finally, since for each reconstruction we are using only 1 atom (which in this case corresponds to a whole glitch), the classification dictionaries' effectiveness will increase with their number of atoms, as can be seen in results of figure \ref{fig:natoms}. From these, we consider 256 atoms to be a good trade-off between accuracy and performance, although sensibly better results could be achieved by using even more.

\subsection{Final test}

So far we have evaluated every configuration using a few testing signals, which was enough to make comparisons. However, for the final configuration almost all glitches were classified correctly, not leaving enough data to evaluate whether the dictionaries are well balanced between them. Therefore we repeat the case study using the testing set of 3000 glitches, with the same parameters as before (summarized in Table~\ref{tab:final_configuration}). In addition, following~\cite{Powell:2015}, for each glitch we now choose a random SNR linearly distributed between 1 and 400.

\begin{table}[!ht]
\caption{\label{tab:final_configuration} Parameter values of the final configuration for the denoising (\textit{den}) and classification (\textit{clas}) dictionaries.}
\begin{indented}
\item[]\begin{tabular}{@{}lllllll}
\br
Dictionary & $\lambda_{\text{learn}}$ & $a_{\text{den}}$ & $n_{\text{den}}$ & $\lambda_{\text{tr}}$ & nonzero & $a_{\text{clas}}$ \\
\mr
sine Gaussian & 0.02 & 512 & 256 & 0.09 & 1 & 256\\
Gaussian & 0.006 & 256 & 128 & 0.09 & 1 & 256\\
Ring-Down & 0.01 & 512 & 256 & 0.09 & 1 & 256\\
\br
\end{tabular}
\end{indented}
\end{table}

\begin{figure}[!ht]
  \centering
  \includegraphics[width=5.5cm]{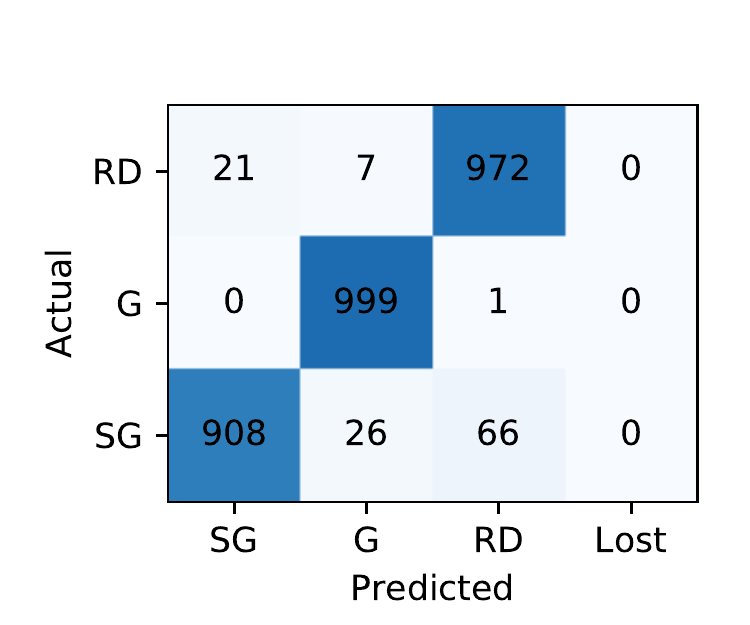}
  \caption{\label{fig:final_results} Confusion matrix of the classification results of the final configuration. Rows correspond to the actual morphology of validation glitches, and columns to the morphology predicted by classification dictionaries.}
\end{figure}

The results of this test are shown in figure~\ref{fig:final_results}. In total, 3000 glitches have been processed, with 2879 (96\%) correctly classified. The Gaussian dictionary recognized all but 1 of its glitches, the one with the lowest SNR, while 92 SG and 28 RD glitches were not correctly classified. The great accuracy of the G dictionary is due to the difference between the Gaussian morphology and the other two. SG and RD glitches can be relatively easy to reproduce using a few Gaussian atoms but not the other way around, because most of their atoms contain more than one oscillation. Hence, SG and RD glitches are more likely to be misclassified.

\begin{figure}[t]
  \centering
  \includegraphics[width=14cm]{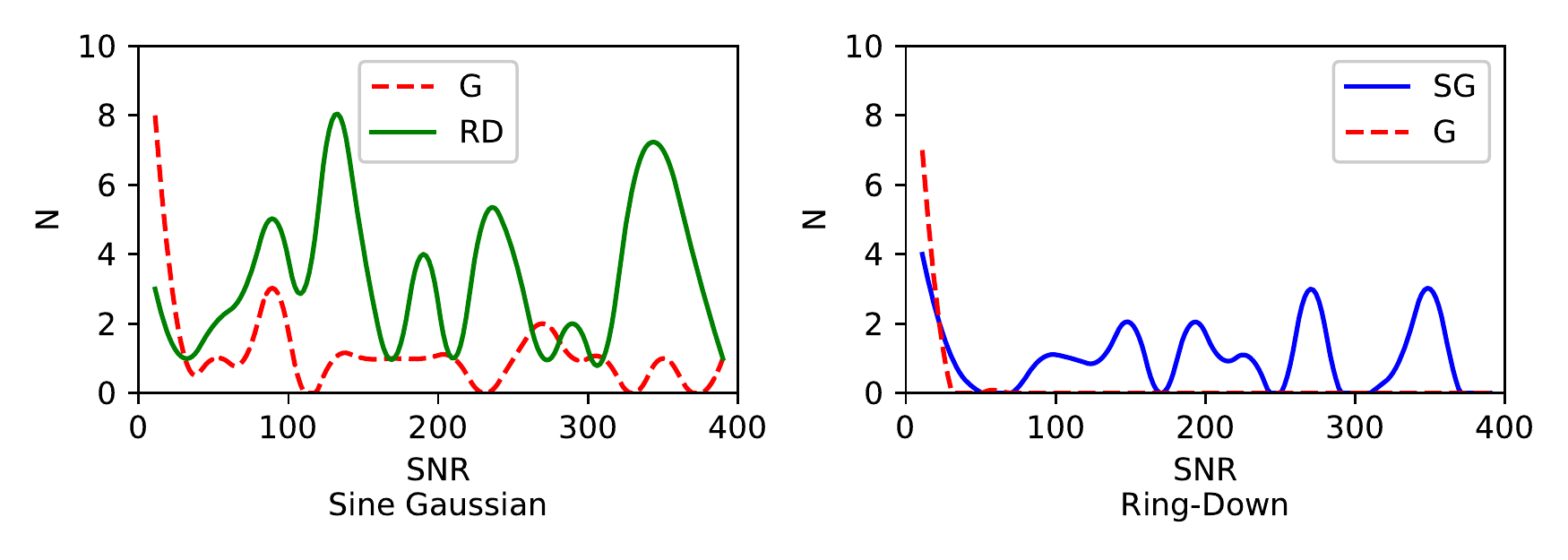}
  \caption{\label{fig:missmatch_snr_distribution} Distribution of misclassified SG glitches (left) and RD glitches (right) as a function of their SNR. Each line represents a (wrongly) predicted morphology.}
\end{figure}

This, however, does not explain the differences found in the number of misclassified glitches (also called false positives) between SG and RD dictionaries. For the sake of understanding the results we analyze how the classification method behaves for misclassified glitches depending on their SNR and frequency. The results of this analysis are shown in figure~\ref{fig:missmatch_snr_distribution}. The left panel corresponds to SG glitches misclassified as G or RD and the right panel to RD glitches misclassified as G or SG. In figure~\ref{fig:missmatch_snr_distribution} we can see roughly the same ratio of false positives within all the range of SNR values, except for the lowest SNR (below 10) where most mismatches occur (as expected). These false positives come mostly from the Gaussian dictionary; at such low SNR both SG and RD dictionaries yield more irregular reconstructions because of noise fluctuations, while Gaussian reconstructions are cleaner, making them more likely to be identified as such. Therefore, the SNR may unbalance the number of false positives only in favor of the Gaussian morphology.

\begin{figure}[!ht]
  \centering
  \includegraphics[width=14cm]{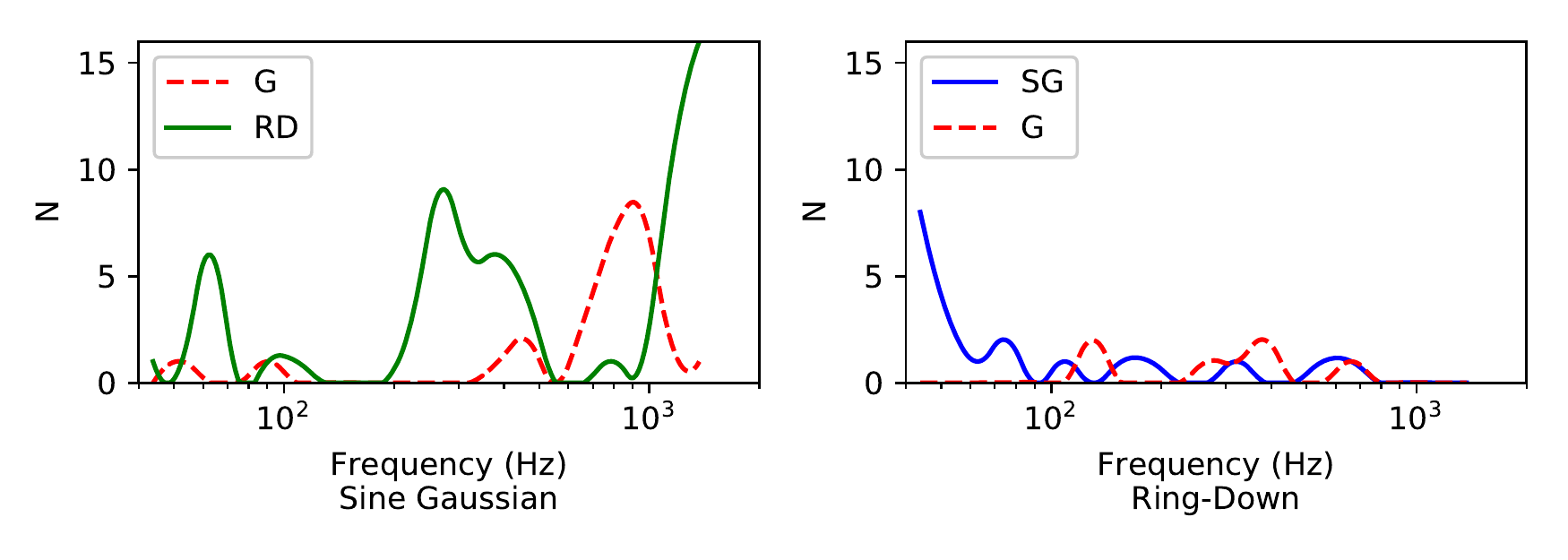}
  \caption{\label{fig:missmatch_frequency_distribution} Distribution of misclassified SG glitches (left) and RD glitches (right) as a function of their frequency. Each line represents a (wrongly) predicted morphology.}
\end{figure}

On the other hand, the results for the distribution in terms of frequency, shown in figure \ref{fig:missmatch_frequency_distribution}, are more inhomogeneous. The left panel indicates that SG glitches are misclassified as G and RD more often at high frequencies, while RD glitches are misclassified mainly as SG at low frequencies. High frequency SG glitches have in general shorter durations (see Eq.~\eqref{eq:glitch_equations}), which combined with noise fluctuations makes them more similar to short Gaussian or RD glitches, explaining the high frequency peaks in the left plot and the greater number of wrongly predicted RD glitches compared to that of SG glitches. At the same time, glitches with lowest frequencies are generally much larger than the atoms' window (about an order of magnitude), so they need to be split in multiple samples, each of them containing usually only a few oscillations. Individually, these windows are almost indistinguishable from both SG and RD's low frequency atoms in terms of morphology, which explains the similar increase in misclassified glitches in both plots at lower frequencies.

In summary, our method was able to classify by morphology 96\% of testing glitches with a wide range of parameters and SNR. It succeeded in recognizing Gaussian glitches, and achieved a good discrimination level between SG and RD glitches, being slightly unbalanced mainly due to their morphological dissimilarities. More precision may be achieved by using more atoms for the untrained dictionaries, or even splitting each denoising dictionary into several ones for different frequency intervals.

\section{Conclusions}
\label{sec:conclusions}

In this paper we have introduced a new method for the classification of transient noise signals (or glitches) by their waveform morphology in advanced GW  interferometers. The method uses learned dictionaries (a supervised machine learning algorithm) for the denoising, and untrained dictionaries for the final sparse reconstruction and classification. To test the accuracy of our method we have used a set of simulated glitches embedded in non-white Gaussian noise, simulating the background noise of advanced LIGO in the proposed broadband configuration.

Using a data set consisting of 3000 glitches divided into three different waveform morphologies with a large range of parameters, the method has shown a 96\% classification accuracy. Furthermore, its performance has been found to barely decrease with the SNR, down to low SNR values ($\sim 10$). Its main limitation appears at extreme frequencies; most of misclassified glitches have the highest or lowest frequencies. In the case of low SNR, it is challenging to discriminate the glitches from the noise and the algorithm tends to misclassify them. Something similar explains what happens for low and high frequencies. As the detector sensitivity is not flat in frequency, those frequencies are more affected by noise than the middle ones. Nevertheless, our study with simulated glitches and Gaussian noise shows that dictionaries are successful at discriminating even the two most similar morphologies of our sample.

There are possible extensions of this study we plan to undertake next. Certainly, the accuracy of our method can be significantly improved by using more untrained atoms in order to expand the model population (especially at extreme frequencies), as the results of figure~\ref{fig:number_atoms} indicate. It may be interesting to study by how much the accuracy could be improved by using several learned dictionaries for each waveform morphology, in order to split the frequency range into smaller intervals. We also plan to implement additional methods to perform the LASSO algorithm more efficiently, since the denoising phase of our current approach is the most computationally expensive. Reconstructing all 3000 glitches, each one inside one patch of 16384 samples, by all three learned dictionaries takes about 36 hours with an AMD Phenom II x4 processor and 12 GB of RAM. In addition, in this work we have assumed all signals to be glitches and their positions to be already known. Our method does not provide a trigger for glitches, therefore a real GW signal or even a pure noise sample could be classified as a glitch. The first case can be avoided by removing coincident signals before the classification. For the latter, it may be worth analyzing whether a previous denoising phase using learned dictionaries with lower sensitivities could be used as a trigger layer before employing our current method.

This work constitutes a preliminary step before assessing the performance of dictionary-learning methods with actual detector glitches. A natural next step is to test our method employing glitches from Advanced LIGO's first observing run (O1) whose waveform morphologies have been already classified~\cite{Zevin:2017}. In a longer time frame we plan to combine learned dictionaries for both glitches and GW signals and try to discriminate if a detector trigger can be classified as a glitch or a signal. This would allow to compute the probabilities of false alarm (identifying a glitch as a true signal) and of losing a signal (either because it has been classified as a glitch or as background noise). Such a study appears necessary to explore the usefulness of dictionary-learning algorithms for GW data analysis.

\ack
Work supported by the Spanish MINECO (grants AYA2015-66899-C2-1-P and MTM2014-56218-C2-2-P), by the Generalitat Valenciana (PROMETEOII-2014-069), and by the European Gravitational Observatory (EGO-DIR-51-2017).

\section*{References}
\bibliographystyle{iopart-num}
\bibliography{paper}

\end{document}